\documentclass{ws-procs}

\newcommand{\ellK}{{\rm K}}

\newcommand{\ellE}{{\rm E}}

\begin{document}


\title{TWO-LOOP COMMUTING CHARGES AND THE STRING/GAUGE DUALITY}

\author{Gleb Arutyunov and Matthias Staudacher}
\address{
Max-Planck-Institut f\"ur Gravitationsphysik\\
     Albert-Einstein-Institut \\
     Am M\"uhlenberg 1, D-14476 Potsdam, Germany\\
agleb@aei.mpg.de, matthias@aei.mpg.de
}


\maketitle

\abstracts{
We briefly review the status quo of the application of
integrable systems techniques to the AdS/CFT correspondence
in the large charge approximation, a rapidly evolving topic.
Intricate string and gauge computations of, respectively,
energies and scaling dimensions agree at the one and two-loop
level, but disagree starting from three loops. To add to this
pattern, we present further computations which demonstrate that for
folded and circular spinning strings the full tower of infinitely
many hidden commuting charges, responsible for the integrability,
also agrees up to {\it two}, but not three, loops.
}


\section{The status quo}

A fresh approach to the AdS/CFT correspondence was initiated in
\cite{bmn}. The proposal involved e.g.~the study of composite operators
in the planar $\mathcal{N}=4$ gauge theory containing as constituents
complex scalar fields $Z$,$\Phi$ of the form
\begin{equation}
\label{bmnop}
\mathop{\mathrm{Tr}} Z^{J_1} \Phi^{J_2}  + \ldots \, ,
\end{equation}
and to consider the situation where $J=J_1+J_2$ becomes large,
while $J_2$ remains small. The dots indicate arbitrary
orderings of the scalars inside the trace.
It was argued that in this limit
the conjectured dual string becomes essentially free, and
the spectrum therefore explicitly calculable. Furthermore, the
string energies turned out to be {\it analytic} at small
BMN coupling
\begin{equation}
\label{lambdaprime}
\lambda'=\frac{\lambda}{J^2} \, ,
\end{equation}
where $\lambda$ is the usual large color 't~Hooft coupling.
The string energy is interpreted on the gauge side as
the anomalous dimension of the corresponding operator eq.(\ref{bmnop}).
Expanding in $\lambda'$ therefore {\it looks} like a perturbative
expansion of the dimension in the dual gauge theory. And indeed
the one- and two-loop predictions for the BMN operators
eq.(\ref{bmnop}) were successfully reproduced in \cite{bmn},\cite{gross}.

The large $J$ limit of \cite{bmn} was given in \cite{gkp}
an alternative interpretation as a semi-classical approximation
to the string sigma model. This way of thinking allows for a
generalization of the BMN limit: One may, more generally, assume
several charges to be large \cite{ft2} (see also \cite{ft0}, and
the closely related earlier work \cite{earlier}). In the case
of our operators in eq.(\ref{bmnop}) one thus assumes
both $J_1$ and $J_2$ to be large. It was then shown \cite{ft2}
that the semi-classical computations of string energies
could be performed exactly, leading to predictions for
the anomalous dimensions of these operators which are rather
intricate as they generically depend in mathematically complicated
ways on both the coupling $\lambda'$ and the ratio
\begin{equation}
\label{alpha}
\alpha=\frac{J_2}{J} \, .
\end{equation}
In subsequent papers it was understood that the reason for
the exact solvability of the semi-classical string motions
may be traced back to {\it integrability} \cite{afrt}.
For a
string propagating on a five-sphere the evolution equations are those
of the O(6) sigma model. In light-cone coordinates $(\xi,\eta)$,
where $\tau=\xi+\eta$ and $\sigma=\xi-\eta$, they read
\begin{equation}
\label{sem}
X_{\xi\eta}+(X_{\xi}\cdot X_{\eta})X=1 \, , ~~~~(X\cdot X)=1 .
\end{equation}
Here $X$ is a six-dimensional vector describing the embedding
of the string world-sheet in $S^5$. The equations are supplemented
by the Virasoro constraints. The simplest solutions correspond to
rigid strings, i.e.~string configurations with a time-independent shape.
Rigid strings are naturally classified in terms of the
Neumann integrable system \cite{afrt},
which is a special one-dimensional reduction of eqs.(\ref{sem}).

The string equations of motion inherit their integrability
from the one of the string sigma model \cite{Pol}.
The underlying reason for the solvability is then
an infinite family of mutually
commuting Pohlmeyer charges $\{\mathcal{Q}_{k}\}$,
where $\{\mathcal{Q}_{2}\}$ denotes the string energy.

How to check these predictions for the $\mathcal{N}=4$ gauge theory?
The problem is a priori difficult, even at one-loop,
due to the complicated mixing between operators differing
by the ordering of the scalars inside the trace. To deal with
these problems, operatorial methods were developed in
\cite{effective}, not necessarily restricted to the planar
case. However, for the planar case, a very important feature
was noticed in \cite{mz}: The one-loop anomalous dimensions may
be obtained by the diagonalization of an {\it integrable}
quantum spin chain. In the example of the operators of
eq.(\ref{bmnop}) the $Z$ and $\Phi$ fields correspond to
``up'' and ``down'' spins.
This integrable structure leads to the
appearance of an infinite set of mutually commuting one-loop
gauge charges $\{Q^{(1)}_{k}\}$, which generate (planar) symmetries
which, mysteriously and yet-to-be-understood, are {\it not}
part of the known symmetries of the superconformal $\mathcal{N}=4$ theory.
With one exception: $\{Q^{(1)}_{2}\}$, which is the
one-loop approximation to the model's dilatation operator.
Excitingly, investigating two-loop corrections, it was
found that the charges continue to commute beyond the one-loop
level \cite{bks}. This led to the conjecture \cite{bks} that
the one-loop charges are actually only the leading order
approximation of coupling constant dependent commuting charges
$\{Q_{k}(\lambda)\}$, and that the exact dilatation
operator $\{Q_{2}(\lambda)\}$ might thus be integrable.

The integrability of the one-loop planar gauge theory for
scalar operators leads to the existence of a Bethe Ansatz for the
anomalous dimensions \cite{mz}. Incidentally,
one can show that this remains true for the
set of all planar one-loop anomalous dimensions \cite{bs},
as the one-loop dilatation operator, not necessarily restricted to
the planar limit, is known for the most general
$\mathcal{N}=4$ composite operator \cite{b1}.
The Bethe Ansatz allows one to quantitatively address the problem
of the large $J_1$,$J_2$ limit of the operators in eq.(\ref{bmnop}),
and to subsequently compare to the semi-classical string results.
And indeed, this programme recently enjoyed considerable success.
Investigating the two simplest types of string motions
corresponding to the operators eq.(\ref{bmnop}) (the {\it folded}
and the {\it circular} string), it was
shown in {\cite{bmsz},\cite{bfst}
that the one-loop string
energies \cite{ft2},\cite{ft4},\cite{afrt} and the gauge theory
anomalous dimensions agree:
$\mathcal{Q}_2^{(1)} \sim Q_2^{(1)}$. For agreement for
other types of motion, see the string \cite{ftfollowup}
and gauge computations \cite{ymfollowup}; for a review,
see also \cite{arkady}.

In \cite{as} it was shown, for folded and circular strings, that
{\it all} one-loop charges $\mathcal{Q}_{2 k}^{(1)} \sim Q_{2 k}^{(1)}$
agree, establishing a direct connection between the respective
integrable structures. This involved applying a B\"acklund
transformation to the string motions, and extracting
the charges from the Bethe Ansatz for the gauge theory. For a beautiful
extension of this approach to the case involving three
complex scalars, see \cite{engquist}.
In \cite{ss} it was shown that the two-loop energies
$\mathcal{Q}_2^{(2)} \sim Q_2^{(2)}$ agree. This required
a generalization of the spin chain picture to a long range chain,
and an extension of the Bethe Ansatz to higher loops\footnote{
The procedure applied in \cite{ss}, using a long range chain
invented by Inozemtsev, accurately diagonalizes
the dilatation operator for the states eq.(\ref{bmnop}) up
to three loops. At four loops, a violation of BMN scaling
was detected. This might either indicate that BMN scaling
is indeed violated in the gauge theory, or alternatively,
that a further integrable long range chain, different from Inozemtsev,
exists. Some evidence for the second scenario comes from the
results of \cite{bes4}. However, this issue is beyond the present
scope.}.

Below we will extend
these matchings by proving that also the two-loop charges
$\mathcal{Q}_{2 k}^{(2)} \sim Q_{2 k}^{(2)}$ coincide.

While the matching is done for the two specific cases
we are convinced that the relations between string and gauge
charges are universal.
As we will see, they are not only independent of the
type of motion, but also of the parameter
$\alpha$ in eq.(\ref{alpha}).
Of course it would be nice to present a general,
solution independent proof. A first important
step into this direction was undertaken, at one loop,
in {\cite{krucz}.

Very recently the most general one- and two-loop solution
of the large $J$ gauge theory scaling dimensions for the
operators eq.(\ref{bmnop}) was obtained by Kazakov, Marshakov, Minahan and
Zarembo  \cite{kmmz}. Furthermore,
a Bethe equation for the classical string sigma model
was derived, and it was argued that the quantum spin chain
Bethe equations may be obtained from it, mapping
the integrable structures up to two loops for the most
general solution.
These results based on the monodromy approach are completely
consistent with our extension of the B\"acklund approach to two
loops described in this work below.

Not all is well, however. Serious problems appear at the
{\it three} loop level. Indeed, the three-loop dilatation operator
for the operators eq.(\ref{bmnop}) had been conjectured
in \cite{bks}, see also \cite{bes4}, and rigorously derived in \cite{dynamic}.
Completely independent confirmation comes from a conceptually,
but not technically related study of plane wave matrix theory at
three loops \cite{klose}.
Strictly speaking, the three loop dilatation generator is
only known up to two unknown constants:
The procedure of \cite{dynamic} allows, to the considered order,
to add to $Q_2^{(3)}$, with arbitrary coefficients,
the second ($Q_2^{(1)})$ and fourth $(Q_4^{(1)})$
one-loop conserved charge. However, both additions would be inconsistent
with the perturbative BMN limit, and therefore certainly also with
the large $J_1$,$J_2$ ``spinning limit''. However, three-loop
integrability is {\it proven} in \cite{dynamic}. Therefore,
the Bethe Ansatz of \cite{ss} applies, but the obtained
three loop string and gauge energies
{\it disagree}, $\mathcal{Q}_2^{(3)} \nsim Q_2^{(3)}$, and the same is
true for all other charges.
Below we will show that the
difference is nevertheless ``small'' in the sense that it may always be
accounted for by a simple non-local combination of charges,
generalizing the ``curious observation'' of \cite{ss}.
An earlier three-loop string-gauge disagreement was reported
for the easier (on the gauge side) case of $1/J$ corrections
to the BMN limit in \cite{callan}.

Let us assume that the AdS/CFT correspondence is not a near-symmetry,
but actually correct. One possible explanation for
the disagreement might be the fact that the parameter
$\lambda'$ in eq.(\ref{lambdaprime})
is not obtained in the same fashion in string and gauge theory.
This possibility was first discussed, to our knowledge, in \cite{ss}.
For an earlier discussion in a slightly different but related
context, see \cite{ksv}.
In the ``semiclassical'' string theory calculations
we take $\lambda$ and $J$ to be large in a coupled fashion,
and form the
finite coupling $\lambda'$. In {\it perturbative} gauge theory,
we first assume $\lambda$ to be small, and subsequently take
$J$ to infinity. The difference in procedure might be expressible as
certain non-perturbative corrections on either side of the
correspondence. Taking them into account will hopefully eventually
show that indeed IIB string theory on the background
$AdS_5 \times S^5$ and $\mathcal{N}=4$ gauge theory are
one and the same thing.

\section{Two-loop string/gauge matching of infinitely many higher
commuting charges}

Let us now show explicitly that the integrable structures
on both sides of the correspondence agree, up to two loops,
in the two distinct cases of the folded and the circular
spinning string.
This will be done by combining the results of \cite{as} and
\cite{ss} in order to show that the full tower of commuting
charges matches up to $\mathcal{O}({\lambda'}^2)$.
In the following we will concentrate on the new, two-loop aspects
of this matching, and refer to \cite{as},\cite{ss} for further background,
technical details, and precise notations.
On the string side the generating function of charges
(resolvent) was found by applying the B\"acklund transformation
to the solutions of the classical string equations of motion,
{\it cf.}~eqs.(\ref{sem}). The result
for the ``nearly improved'' string resolvent reads, in both
cases,
\begin{equation}
\label{sgf}
\tilde{\mathcal{E}}(\mu)=\frac{4\mu^3}{\pi}\frac{\sqrt{(1-z)(1-tz)}}{\sqrt{z}}
~\Pi(tz,t).
\end{equation}
Here $\mu$ is the string spectral parameter, $t$ is the string modulus, and
the function $\Pi(t z,t)$ is the complete
elliptic integral of the third kind. The auxiliary parameter $z$
depends on $\mu$, $t$, and $\mathcal{J}=J/\sqrt{\lambda}$,
i.e.~the charge in units of the string tension.
In order to expand this expression, it is convenient to use the form
\begin{equation}
\label{backlund}
\tilde{\mathcal{Q}}(\varphi,\lambda')=
\frac{1}{2}-\frac{1}{2}~\mu^{-2}~\frac{\tilde{\mathcal{E}}(\mu)}{\mathcal{J}}=
\frac{1}{2}
-2 \varphi \frac{\sqrt{\mp (1-z)(1-tz)}}{\sqrt{z}}~\Pi(tz,t) \; ,
\end{equation}
The string resolvent then becomes a function of the rescaled
spectral parameter $\varphi$
and the BMN coupling constant $\lambda'$,
\begin{equation}
\varphi^2=\mp\frac{\mu^2}{\pi^2 \mathcal{J}^2}
\qquad {\rm and} \qquad
\lambda'=\frac{1}{\mathcal{J}^2} \; .
\end{equation}
Here and in the following the sign $\mp$ depends on the type of string motion
(folded: upper sign, circular: lower sign), which is entirely due to our
conventions. The modulus $t=t(\lambda';\alpha)$ depends on the
coupling constant $\lambda'$, and also
encodes information on the filling fraction $\alpha=J_2/J$,
{\it cf.}~eqs.(\ref{lagrange1}),(\ref{alpha1}) and
eqs.(\ref{lagrange2}),(\ref{alpha2}).
The auxiliary parameter $z=z(\varphi,\lambda';\alpha)$ also slightly
differs in the two considered cases.
It is determined by eqs.~(\ref{zfolded}),(\ref{zcircular}) below.

This ``nearly improved'' string resolvent generates a set of string charges
$\{\tilde{\mathcal{Q}}_{2 k}(\lambda')\}$
\begin{equation}
\label{stringcharges}
\tilde{\mathcal{Q}}(\varphi,\lambda')=
\sum_{k=0}^{\infty} \tilde{\mathcal{Q}}_{2 k}(\lambda')~\varphi^{2 k},
\end{equation}
such that the lowest (zeroth) charge is linearly related to the string energy
$\mathcal{Q}_2(\lambda')=\mathcal{E}/\mathcal{J}$
by $\tilde{\mathcal{Q}}_0(\lambda')=\frac{1}{2}(1-\mathcal{Q}_2(\lambda'))$.
Of course we can alternatively, or in addition,
expand in the coupling constant $\lambda'$:
\begin{equation}
\tilde{\mathcal{Q}}(\varphi,\lambda')=
\sum_{n=0}^{\infty} \tilde{\mathcal{Q}}^{(n+1)}({\varphi})~{\lambda'}^n=
\sum_{n,k=0}^{\infty} \tilde{\mathcal{Q}}^{(n+1)}_{2 k}~{\lambda'}^n~\varphi^{2 k}.
\end{equation}

On the gauge side the resolvent for a set of gauge charges
$\{\bar{Q}_{2 k}(\lambda')\}$ was also found perturbatively, up to
three loops, by using a Bethe Ansatz for a long range spin chain
originally invented by Inozemtsev \cite{inozemtsev},\cite{ino02}:
\begin{equation}
H(\varphi,\lambda')=
\pm \sum_{k=1}^{\infty}
\bar{Q}_{2 k}(\lambda')~\varphi^{2 k} \; .
\end{equation}
Here $\varphi$ is a gauge spectral parameter which is obtained from the scaled
rapidity of the Inozemtsev-Bethe Ansatz.
Expanding in
$\lambda'$, we have
\begin{equation}
H(\varphi,\lambda')=
\pm
\sum_{n=0}^2 \bar{Q}^{(n+1)}({\varphi})~{\lambda'}^n=
\pm \sum_{n=0}^2 \sum_{k=1}^{\infty}
\bar{Q}^{(n+1)}_{2 k}~{\lambda'}^n~\varphi^{2 k}.
\end{equation}
It is important to note that the charges $\bar{Q}^{(n)}_{2 k}$ are
not the correct ``observables'' of the spin chain.
E.g.~the three-loop gauge anomalous dimension is obtained by the following
specific linear combination:
\begin{equation}
\label{thermoenergy}
Q_2(\lambda')=1 \mp \frac{\lambda'}{4 \pi^2}~\bar{Q}_2(\lambda')
-\frac{3 {\lambda'}^2}{64 \pi^4}~\bar{Q}_4(\lambda') \mp
\frac{5 {\lambda'}^3}{512 \pi^6}~\bar{Q}_6(\lambda').
\end{equation}
Likewise, while in the spin chain any linear combination of local charges
leads again to a set of local charges, in the gauge theory these
linear superpositions are essentially fixed: At one loop, the
charges are required to obey BMN scaling \cite{as}, and loop
corrections are then determined by quantum field theory. The proper
three-loop charges will thus look like ($k>0$)
\begin{equation}
Q_{2 k}(\lambda')=\bar{Q}_{2 k}(\lambda')
\mp e_{k,1} \lambda' ~\bar{Q}_{2 k+2}(\lambda')
+e_{k,2} {\lambda'}^2 ~\bar{Q}_{2 k+4}(\lambda') \; .
\end{equation}

While it is certainly possible to find these (universal) numbers
$e_{k,1}$, $e_{k,2}$, it is easily seen that we do not require
them for the present purposes of two-loop matching: All we need to show\footnote{
This also relieves us from applying the procedure of ``full improvement''
to the string resolvent eq.(\ref{sgf}), as discussed in \cite{as}, p.~24.
At any rate, the latter would only assure us of the correct leading BMN scaling
behavior of the charges; one can show that the obtained charges
do {\it not} agree with the ``proper'' gauge charges at the two-loop level.
Thus a further linear, upper triangle redefinition would be
required.}
is that a linear map from the set of ``nearly improved'' string charges
$\{\tilde{\mathcal{Q}}_{2 k}(\lambda')\}$
to the set of gauge charges $\{\bar{Q}_{2 k}(\lambda')\}$
exists to leading and next-to-leading order in $\lambda'$!
By explicit inspection of the first few two-loop string and gauge charges,
one finds that ($k \geq 0$, and $\bar{Q}^{(2)}_{0}\equiv 0$)
\begin{equation}
\label{prop1}
\tilde{\mathcal{Q}}^{(2)}_{2 k}
=\pm \bar{Q}^{(2)}_{2 k}
\pm \frac{1}{ 8 \pi^2}(2 k+1)~\bar{Q}^{(1)}_{2 k+2} \, .
\end{equation}
This leads to the following proposition
\begin{equation}
\label{mainresult}
\tilde{\mathcal{Q}}(\varphi,\lambda')=
\bar{Q}^{(1)}(\varphi) \pm \lambda' \left[
\bar{Q}^{(2)}(\varphi)+\frac{1}{8 \pi^2} \frac{\partial}{\partial \varphi}
\frac{1}{\varphi} ~\bar{Q}^{(1)}(\varphi) \right]
+ \mathcal{O}({\lambda'}^2)
\end{equation}
It can be written in an elegant fashion as follows:
\begin{equation}
\label{el}
\tilde{\mathcal{Q}}(\varphi,\lambda')=
\Big(1\mp\frac{\lambda'}{8\pi^2\varphi^2}  \Big)
H\Big(\varphi\pm\frac{\lambda'}{8\pi^2\varphi} ,\lambda' \Big)+ \mathcal{O}({\lambda'}^2)\, .
\end{equation}
This is our main new result,
and will be proven in the next sections for the case of the
folded and circular string. It proves our assertion of
the two-loop matching of string and gauge charges.
Although derived starting from particular solutions,
we believe that (\ref{el}) is universal in the sense that
it does not depend on specific solutions\footnote{The sign flips are
entirely due to a
slight difference in the normalization of gauge theory charges
for the folded and circular cases, {\it cf.}
\cite{as}.
Defining the charges in an identical fashion in both cases
removes this difference.}, and that it expresses the general
matching of string/gauge integrable structures at two loops.

Now we discuss the three-loop case. As was already noticed in \cite{ss}
the gauge/string energies disagree starting from three loops.
It is however remarkable that the disagreement between the whole towers
of gauge/string commuting charges admits a uniform description.

First, we expand eq.(\ref{backlund}) up to ${\lambda'}^2$ and
identify the term $\tilde{\mathcal{Q}}^{(3)}(\varphi)$. Then
we perform the Gauss-Landen transformation (\ref{GLF}) to bring this term
to the gauge theory frame.
Our two-loop matching formula (\ref{prop1}) would seem to suggest that
the matching of the three-loop gauge and string resolvents
should again be given by a linear relation
\begin{equation}
\label{prop}
\tilde{\mathcal{Q}}^{(3)}_{2k}\stackrel{?}{=}
\bar{Q}^{(3)}_{2k}+
\alpha_{k,1}~\bar{Q}^{(2)}_{2k+2}+
\alpha_{k,2}~\bar{Q}^{(1)}_{2k+4}\, .
\end{equation}
Working out explicitly the first few Inozemtsev charges
one may verify that this proposal does not work, since
the coefficients $\alpha_{k,1}$ and $\alpha_{k,2}$
then appear to be functions of the modular parameter $q_0$.
Thus, linear combinations of Inozemtsev charges with constant
coefficients cannot reproduce the string result. This motivates us
to extend the set of charges by adding also their products.
By trial and error we found the following remarkable formula
($k\geq0$, and $\bar{Q}^{(2)}_{0}\equiv \bar{Q}^{(3)}_{0}\equiv 0$)
\begin{eqnarray}
\nonumber
\tilde{\mathcal{Q}}^{(3)}_{2k}&=&
\bar{Q}^{(3)}_{2k}
+\frac{2k+1}{8\pi^2}~\bar{Q}^{(2)}_{2k+2}
+\frac{(2k+1)(2k+3)}{128\pi^4}~\bar{Q}^{(1)}_{2k+4} \\
&&~~~~
-\frac{1}{8\pi^2}~\bar{Q}^{(1)}_{2}~\bar{Q}^{(2)}_{2k}\,
\end{eqnarray}
which describes the relation between the gauge and string towers of commuting
charges up to and including three loops.

\subsection{Folded string}

The classical motion of the folded string leads to the following
parametric result for the all-loop string energy $\mathcal{Q}_2$:
\begin{eqnarray}
\label{lagrange1}
\lambda'&=&\frac{\pi^2}{4 \ellK(t_0)^2}
\left[ \left( \frac{\ellK(t_0)-\ellE(t_0)}{\ellK(t)-\ellE(t)} \right)^2
-\left(\frac{\ellE(t_0)}{\ellE(t)} \right)^2 \right], \\
\mathcal{Q}_2&=&\frac{\ellK(t)}{\ellK(t_0)} \sqrt{
(1-t)\left(\frac{\ellE(t_0)}{\ellE(t)} \right)^2+
t \left( \frac{\ellK(t_0)-\ellE(t_0)}{\ellK(t)-\ellE(t)} \right)^2 }.
\nonumber
\end{eqnarray}
Here the modulus $t$ contains the information about the coupling
$\lambda'$. In addition, its constant piece $t_0$ parametrizes the
``filling fraction''
\begin{equation}
\label{alpha1}
\alpha=1-\frac{\ellE(t_0)}{\ellK(t_0)} \; .
\end{equation}
The charges are found by the B\"acklund transformation; the final result
for the string resolvent is given in eq.(\ref{backlund}), where the
``auxiliary'' and true spectral parameters $z,\varphi$ are related
through
\begin{equation}
\label{zfolded}
\varphi^2
+\left(\frac{1}{4 \ellK(t_0)}
\frac{\ellE(t_0)}{\ellE(t)}\right)^2\frac{z(1-tz)}{1-z}
+\frac{\lambda'}{4\pi^2} t z =0\, .
\end{equation}
Expanding this in $\lambda'$ one finds
\begin{eqnarray}
\label{fexp}
\tilde{\mathcal{Q}}(\varphi,\lambda')=\tilde{\mathcal{Q}}^{(1)}(\varphi)
+\lambda'\tilde{\mathcal{Q}}^{(2)}(\varphi)+
\mathcal{O}({\lambda'}^2) \, .
\end{eqnarray}
Here the leading piece is
\begin{eqnarray}
\label{foldedexp}
&&\tilde{\mathcal{Q}}^{(1)}(\varphi)=
\frac{1}{2}
-2 \varphi \frac{\sqrt{(1-z_0)(t_0 z_0-1)}}{\sqrt{z_0}}~\Pi(t_0 z_0,t_0)\, .
\end{eqnarray}
where
\begin{equation}
\varphi^2
+\left(\frac{1}{4 \ellK(t_0)} \right)^2\frac{z_0(1-t_0 z_0)}{1-z_0} =0\, .
\end{equation}
The next term $\tilde{\mathcal{Q}}^{(2)}(\varphi)$ is rather
complicated and we will not write it out.

The Bethe Ansatz for the spin chain describing two-loop gauge theory
in the large $J$ limit quite generally leads to a singular integral equation.
In the present case this equation is of elliptic type, and closely related
to the one appearing in the so-called O($N$) matrix model, see e.g.~\cite{KS}.
Its solution leads to the following generating functions for the one-loop
charges $\bar{Q}^{(1)}_{2 k}$
\begin{equation}
\label{fc}
\bar{Q}^{(1)}(\varphi)=
\frac{1}{4}
-\frac{a^2_0}{b_0} \sqrt{\frac{b_0^2-\varphi^2}{a_0^2-\varphi^2}}
~\Pi\left(-q_0\frac{\varphi^2}{a_0^2-\varphi^2},q_0\right)
\end{equation}
and two-loop charges $\bar{Q}^{(2)}_{2 k}$
\begin{equation}
\bar{Q}^{(2)}(\varphi)=\frac{1}{32\pi^2\varphi^2}\left(
1-\frac{b_0}{2a_0}\sqrt{\frac{a_0^2-\varphi^2}{b_0^2-\varphi^2}}
-\frac{a_0}{2b_0}\sqrt{\frac{b_0^2-\varphi^2}{a_0^2-\varphi^2}}
\right) \, .
\end{equation}
Here the natural modulus $q_0$ parametrizes the filling fraction through
\begin{equation}
\alpha=
\frac{1}{2}-\frac{1}{2\sqrt{1-q_0}}\,\frac{\ellE(q_0)}{\ellK(q_0)} \, .
\end{equation}
while the leading-order rapidity boundaries $a_0,b_0$ are given by
\begin{equation}
a_0=\frac{1}{4 \ellK(q_0)},
\qquad
b_0=\frac{1}{4 \sqrt{1-q_0}~\ellK(q_0)}\, .
\end{equation}
It is now straightforward to calculate the expression
\begin{equation}
\label{gc}
\bar{Q}^{(2)}(\varphi)+\frac{1}{8 \pi^2} \frac{\partial}{\partial \varphi}
\frac{1}{\varphi} ~\bar{Q}^{(1)}(\varphi) =
\frac{a_0^2+b_0^2-8a_0b_0^2\ellE(q_0)}
{64\pi^2a_0b_0\sqrt{(a_0^2-\varphi^2)(b_0^2-\varphi^2)}}\, .
\end{equation}
As was shown in \cite{as}
after a Gauss-Landen transformation of the moduli
\begin{equation}
\label{GLF}
t_0=-\frac{(1-\sqrt{1-q_0})^2}{4\sqrt{1-q_0}}\, ,
\end{equation}
eq.(\ref{foldedexp}) becomes identical to eq.(\ref{fc}). Now we
observe that the same transformation nicely turns the rather
complicated expression $\tilde{\mathcal{Q}}^{(2)}(\varphi)$ in
eq.(\ref{fexp}) into the r.h.s. of eq.(\ref{gc}), proving our
main assertion eq.(\ref{mainresult}).

\subsection{Circular string}
The analysis of the circular string is very similar to one for the
folded case. The BMN coupling constant $\lambda'$ and the string
energy $\mathcal{Q}_2$ are given by
\begin{eqnarray}
\label{lagrange2}
\lambda'&=&\frac{\pi^2 t}{4 t_0^2 \ellK(t_0)^2}
\left[ \left( \frac{\ellK(t_0)-\ellE(t_0)}{\ellK(t)-\ellE(t)} \right)^2
-\left(\frac{\ellE(t_0)-(1-t_0)\ellK(t_0)}{\ellE(t)-(1-t)\ellK(t)}
\right)^2 \right], \\
\mathcal{Q}_2&=&\frac{t \ellK(t)}{t_0 \ellK(t_0)} \sqrt{
\frac{1}{t}\left( \frac{\ellK(t_0)-\ellE(t_0)}{\ellK(t)-\ellE(t)} \right)^2-
\frac{1-t}{t}
\left(\frac{\ellE(t_0)-(1-t_0)\ellK(t_0)}{\ellE(t)-(1-t)\ellK(t)}
\right)^2 },
\nonumber
\end{eqnarray}
where the parameter $t_0$ is determined via the filling fraction
as follows
\begin{equation}
\label{alpha2}
\alpha=1-\frac{1}{t_0} + \frac{1}{t_0}~\frac{\ellE(t_0)}{\ellK(t_0)}.
\end{equation}
The string spectral parameters $z$ and $\varphi$ are related
through
\begin{equation}
\label{zcircular}
\varphi^2- \frac{t^2}{t_0^2}
\left(\frac{1}{4 \ellK(t_0)} \frac{\ellK(t_0)-\ellE(t_0)}{\ellK(t)-\ellE(t)}\right)^2
\frac{z(1-z)}{1-t z}
-\frac{\lambda'}{4\pi^2} \frac{(1-t) z}{1-tz}=0 \, .
\end{equation}
On the gauge theory side the configuration of Bethe roots
corresponding to the circular string is described by two cuts on
the imaginary axis: $ic<i\varphi<id$ and $-id<i\varphi<-ic$, with a constant
condensate in between. For the one-loop problem we denote the
endpoints of the cut as $c_0$ and $d_0$ and introduce the gauge
theory modulus as $r_0=\frac{c_0^2}{d_0^2}$, where
$c_0=\frac{1}{8\ellK(r_0)}$. The modulus $r_0$ is related to the
filling fraction $\alpha$ as follows
$\alpha=\frac{1}{2}-\frac{1}{2\sqrt{r_0}}+\frac{1}{2\sqrt{r_0}}\frac{\ellE(r_0)}{\ellK(r_0)}$.
Then the one-loop resolvent is
\begin{equation}
\label{1loopc}
\bar{Q}^{(1)}(\varphi)=\frac{1}{4} -\frac{2}{d_0}
\sqrt{(d_0^2-\varphi^2)(c_0^2-\varphi^2)}~ \Pi
\left(\frac{\varphi^2}{d_0^2},r_0\right).
\end{equation}
The two-loop correction found from the Inozemtsev-Bethe Ansatz
reads
\begin{equation}
\bar{Q}^{(2)}(\varphi)=\frac{1}{32\pi^2\varphi^2}\left(
1-\frac{d_0}{2c_0}\sqrt{\frac{c_0^2-\varphi^2}{d_0^2-\varphi^2}}
-\frac{c_0}{2d_0}\sqrt{\frac{d_0^2-\varphi^2}{c_0^2-\varphi^2}}
\right)\, .
\end{equation}
Using the two formulae above one finds
\begin{equation}
\label{ccor} \bar{Q}^{(2)}(\varphi)+\frac{1}{8 \pi^2}
\frac{\partial}{\partial \varphi} \frac{1}{\varphi}
~\bar{Q}^{(1)}(\varphi)
=\frac{c_0^2-d_0^2+16c_0d_0^2\ellE(r_0)}{64\pi^2c_0d_0\sqrt{(c_0^2-\varphi^2)(d_0^2-\varphi^2)}}
\end{equation}
Expanding in $\lambda'$ the resolvent (\ref{backlund})
corresponding to the circular string one identifies the leading
and the subleading terms, which are
$\tilde{\mathcal{Q}}^{(1)}(\varphi)$ and
$\tilde{\mathcal{Q}}^{(2)}(\varphi)$ respectively. Quite
remarkably, the Gauss-Landen transformation
\begin{equation}
t_0=-\frac{4\sqrt{r_0}}{(1-\sqrt{r_0})^2}
\end{equation}
transforms these expressions into eqs.(\ref{1loopc}) and
(\ref{ccor}).



\section*{Acknowledgments}

We would like to thank Niklas Beisert for important discussions.
M.S.~thanks Vladimir Dobrev and all other organizers
of the {\it Fifth International Workshop on Lie Theory and
its Applications in Physics} in Varna (Bulgaria) in June 2003
for an inspiring conference, and their warm hospitality.










\end{document}